\documentclass[12pt]{iopart}
\usepackage{iopams}
\usepackage{amsfonts,amssymb,latexsym}

\newcommand{\beq}{\begin{equation}}
\newcommand{\eeq}{\end{equation}}
\renewcommand{\Im}{\mathop{\mathrm{Im}}}

\begin{document}

\noindent{version of 
      2 April 2015}

 \title[Vacuum energy  of a massive scalar field]
 {Vacuum energy density and pressure of a massive scalar field}

 \author{Fernando Daniel Mera$^1$ and S A Fulling$^2$}

 \address{$^1$ Department of Physics, Northeastern University,
 Boston, MA 02115 USA} 

\address{$^2$ Departments of Mathematics and Physics, Texas A\&M 
University,  College Station, TX 77843-3368 USA} 

\begin{abstract}
With a view toward application of the Pauli--Villars 
regularization method to the Casimir energy of boundaries,
we calculate the expectation values of the 
components of the stress tensor of a
confined massive field in $1+1$ space-time dimensions. 
Previous papers by Hays and Fulling are bridged
and generalized.
The Green function for the time-independent Schr\"odinger 
equation 
is constructed from the Green function for the whole line by the 
method of images; 
equivalently, the one-dimensional
system is solved exactly in terms of closed classical paths and 
periodic orbits.  
Terms in the energy density and in the 
eigenvalue density attributable to the two boundaries 
individually and those attributable to the confinement of the 
field to a finite interval are distinguished so that their 
physical origins are clear.
Then the pressure is found similarly from the cylinder kernel,
the Green function associated most directly with an exponential 
frequency cutoff of the Fourier mode expansion.
Finally, we discuss how the theory could be  rendered finite  by 
the Pauli--Villars method.

\end{abstract}


\maketitle

\section{Introduction}

The Casimir energy \cite{Milton,BKMM,LosAlamos}
of a massive scalar field in two space-time 
dimensions, despite the seeming simplicity of the model, has not 
been completely studied.  The 1979 paper of Hays \cite{Hays} calculated 
the energy and the force but did not look at the local energy 
density, a subject of much interest today.  
The more recent paper of Fulling 
\cite{SAF} treated the energy density for a massless scalar field 
from a 
viewpoint of spectral theory and asymptotics, but  did not 
consider the massive field.  
Neither paper calculated pressure directly.
The present article  generalizes
  the works \cite{Hays} and \cite{SAF} and uses methods from 
each.

The primary reason for studying massive fields in this context is 
to be able to conduct a Pauli--Villars regularization 
\cite{PV,Z,stre,BeDu,ans} (see \ref{hist}).
It has become clear \cite{EFM} 
that the traditional ultraviolet cutoff produces unphysical 
results, dependent on the direction of 
``point-splitting''\negthinspace, 
for the counterterms in energy density and pressure near 
perfectly reflecting boundaries;
this development casts some doubt on the claim that such 
approaches to divergences are more ``physical'' than the analytic 
ones (dimensional or zeta).
 The Pauli--Villars method (which occupies a place somewhere 
between the analytic and the cutoff  methods)  
preserves Lorentz invariance, 
and hence one hopes that it will avoid this problem.  
A serious implementation of this strategy requires 
calculations in 
four space-time dimensions, which are deferred to future work,
but here we give it a test drive.
The previous applications of the method that are 
most pertinent to our problem are those to gravitational 
backgrounds, and we review the relevant literature in \ref{hist}.

In \sref{vacuum} the local energy densities 
 $E_\mathrm{Weyl}(t)$,  
$E_\mathrm{per}(t)$, and $E_\mathrm{bdry}(x,t)$,
related respectively to zero-length, periodic, and closed 
reflected classical paths,
are expressed in terms of Macdonald functions. 
(Here $t$ is a temporary regularization parameter.)
These are  expanded in various limits in \sref{lims}.
As expected, the $m\to0$ limit reproduces the known theory of the 
massless field; the $t\to0$ and $m\to\infty$ limits provide 
needed input into the Pauli--Villars construction.
 \Sref{totalenergy} deals 
with the (regularized) total energy and its nontrivial relation 
to the nonconstant density term, 
$E_\mathrm{bdry}(x,t)$.
\Sref{eigen} deals with the eigenvalue density and 
counting function.
\Sref{pressure} and \sref{morepressure}
use the cylinder-kernel method pioneered by Hays \cite{BH,Hays}
to find the various contributions to the expectation value of
the pressure;  \sref{betaterms} presents
 the dependence on the
parameter $\xi$ that labels different possible gravitational 
couplings.
Finally, \sref{renorm} applies the Pauli--Villars procedure.

The key results 
of the paper are the formulas \eref{EWeyl}, \eref{Eper0}, and 
\eref{Ebdry0} for energy density; 
\eref{pWeyl}, \eref{pper0}, and \eref{bdryp} for pressure; and 
\eref{Ebdrygenbeta} for the 
conformal correction to the energy density (the correction to the 
pressure being zero).

\section{Vacuum energy density from closed and periodic %
orbits}\label{vacuum}
We consider a finite interval with either a Dirichlet or a 
Neumann boundary condition at each end,
following the notation of \cite{SAF},
which allows 
the two boundary conditions to be treated simultaneously.
 Thus $H = -\frac{ d^2}{d 
x^2} + m^2$ acts in $L^2(0,L)$ on the domain defined by
\begin{equation} u^{ (1-l ) } (0) = 0, \qquad u^{ (1-r ) } (L) = 
0, \qquad l, r \in \{0,1 \}. \end{equation}
The superscript is the number of derivatives in the 
boundary condition.  Thus $l=0$ means that the left endpoint is 
Neumann, etc.
In nonrelativistic terms we are solving a Schr\"{o}dinger 
equation with potential $V= m^2$. 
 The Green function can be constructed 
from $G_{\infty}\,$,
the Green function on the whole real line,
 by the method of images:
\beq\label{Greenf}
\fl{
\eqalign{
G(\omega^2, x, y) & = G_{\infty} (y) + (-1)^{l} G_{\infty} (-y) + 
(-1)^{r} G_{\infty} (2 L - y ) + (-1)^{l+r} G_{\infty} ( 2 L + y) 
\\ &{}\quad + (-1 )^{l+r } G_{\infty} (-2 L + y) + (-1)^{2 l + r} 
G_{\infty} (-2 L -y )
 \\ &{}\quad + (-1)^{l + 2r} G_{\infty} (4 L - y ) 
+ (-1)^{2 l + 2r } G_{\infty} (4 L + y) + \cdots
 \\ &= \sum_{n = 
0}^{\infty} (-1 )^{n (l + r ) } G_{\infty} (- 2 n L + y) + 
\sum_{n = 0 }^{\infty} (-1 )^{l + n ( l +r ) } G_{\infty} ( - 2 n 
L - y) \\ &{}\quad + \sum_{n = 1}^{\infty} (- 1 )^{- l + n( l + r 
) } 
G_{\infty} (2 n L - y) + \sum_{n = 1 }^{\infty} (-1 )^{n (l + r 
)} G_{\infty} (2 n L + y).
}
}
\eeq
(Here and occasionally elsewhere we suppress some function
arguments to avoid clutter.)
The only difference from \cite{SAF} is that in $G_\infty$
the energy parameter $\lambda$ must  be replaced by 
$\lambda-m^2$.
Thus, many formulas in \cite{SAF} remain valid if we replace
$\omega$ ($\equiv \sqrt{\lambda}$) by
\begin{equation}\kappa \equiv \sqrt{\omega^2 - m^2} \quad
\mbox{(hence $\omega \, d\omega = \kappa \, d\kappa$)},
\label{omegatokappa}\end{equation}
and the basic Green function is 
 \begin{equation}
G_{\infty}(\omega^2, x, y)=  \frac{ i }{2 \kappa } e^{ i \kappa |x-y|}.
\end{equation}
It is easy to
check from first principles that this new $G_{\infty}$,
and hence $G$, 
satisfy the right equation,
\begin{equation}
- \frac{ \partial^2 G}{  \partial x^2} - \kappa^2 G =
 (H_{x} - \omega^2 ) G( \omega^2 , x, y)= \delta (x-y).
\end{equation}

The spectral densities in terms of $\lambda$ for this
 problem are the same as in \cite{SAF} except for the shift
  of the argument variable $\lambda$ by $-m^2$. 
This is exactly to be expected, because we know that adding a constant
 to the potential in the Schr\"{o}dinger equation merely adds that 
constant to all the energies. 
Note that only values of $\omega \geq m$ need to be 
considered, because we know that $H$ has no spectrum below $m$. 
This even comes out automatically in the formalism, because 
if $\kappa$ is imaginary, then the imaginary part of 
$G_{\infty}$ is zero and doesn't contribute to the density 
of states.
When we go to the variable $\omega$ the situation is slighly more 
complicated: $\kappa$ is not just $\omega$ minus a constant,
and that is where some interesting new behavior arises.

The density of eigenvalues is given in terms of the Green function by 
\beq
  \sum_{j}^{\infty}  \delta ( \lambda_{j}  - \lambda ) 
   = \frac{  1 }{ \pi }   \int_0^L   d x  \,
[ \Im \,  G ( \lambda  + i \epsilon,x,y )  ]_{x = y }\,. 
\eeq
It is more convenient to work  with the density with respect to 
$\omega  = \sqrt{ \lambda}$, 
which carries an additional factor $2 \omega$. Then
\beq
\rho (\omega) \,d\omega
 = \frac {2 \omega}{\pi} \, d\omega \int_0^L dx\,
  \Im  \, G ( \omega^2,x,x)  
  =   \frac{2\kappa }{  \pi   }  
\, d\kappa \int_0^L dx\, \Im  \, G ( \omega^2,x,x) .
\eeq
We have by definition 
\beq
\rho(\omega)= \int_0^L dx\,\sigma(\omega,x), \quad
\sigma(\omega,x)=\frac{2\omega}\pi \, \Im G(\omega^2,x,x),
\eeq
and hence
\beq \label{sigmas}
\fl{
\eqalign{ \frac{\pi \kappa }{ \omega }\, \sigma (\omega,x) & =
2 \kappa \hbox{ Im } G(\omega^2 , x, x) \\ &= \sum_{n = 0}^{ 
\infty} (- 1)^{n (l + r ) } \cos (2 \kappa n L ) + \sum_{n = 
0}^{\infty} (-1)^{l + n(l + r) } \cos (2 \kappa (n L+ x ) )\\ 
&{}\quad + 
\sum_{n = 1 }^{\infty} (- 1)^{- l + n ( l + r ) } \cos (2 \kappa 
(n L - x) ) + \sum_{n = 1}^{\infty } (- 1)^{ n (l + r ) } \cos (2 
\kappa n L) \\ & = 1 + 2 \sum_{n = 1}^{\infty} (- 1)^{n (l + r) } 
\cos (2 \kappa n L) + \sum_{n = - \infty}^{\infty} (- 1)^{l + n 
(l + r) } \cos (2 \kappa (x + n L))\\ & \equiv  \frac{\pi \kappa 
}{ \omega } (\sigma_\mathrm{Weyl} + \sigma_\mathrm{per} + 
\sigma_\mathrm{bdry} 
).
}
}
\eeq

The paths connecting $x$ to 
an image charge in \eref{Greenf} can be folded back into the 
original 
interval as paths connecting $x$ to $y$ after some number of 
reflections from the endpoints.  
In \eref{sigmas} these paths connect $x$ 
to itself.  The first term, coming from a path of zero length, 
provides the bulk spectral density of Weyl's famous theorem.  
Paths with an even number of reflections are periodic and provide 
a spatially homogeneous Casimir energy.  Terms with an odd number 
of reflections ``bounce'' off one of the boundaries and yield 
energy distributions somewhat concentrated there.

The 
stress tensor of a scalar field contains a free parameter, $\xi$, 
reflecting an ambiguity in its coupling to the gravitational 
field.  The relevant formulas are reviewed in \sref{pressure} and 
appendix~C.
Until \sref{pressure} we confine attention to the value 
$\xi=\frac14$, for which the 
energy expressions are maximally simple.
In particular, the contribution of the space 
derivatives to the energy density is identical to that of the 
time derivatives, so we 
can write (following \cite{SAF})
\begin{eqnarray}
\langle T_{00} (t, x)\rangle &\equiv E(t,x ) = - \frac{ 1 }{2 } 
\frac{  d }{ d  t } 
\int_{0}^{\infty}  \sigma (\omega,x ) e^{- \omega t} \, d\omega 
\nonumber\\&
\equiv E_\mathrm{Weyl}  (t)  + E_\mathrm{per}  (t)  + 
E_\mathrm{bdry}  (t, x)  . 
\end{eqnarray}
Here $t$ is an ultraviolet cutoff parameter, which can be 
related by a Wick rotation to a difference of physical time 
coordinates.

From \cite[(3.914.1)]{GR},
\begin{equation}
\int_{0}^{\infty}   e^{ - t \, \sqrt{   m^2 + \kappa^2  }  } 
 \cos ( 2 n L  \kappa)   \, d \kappa = 
\frac{  m t  }{ \sqrt[   ]{ t^2 +  (2n L)^2 }  }
 K_{1}  ( m   \, \sqrt[]{  t^2 +  (2n L)^2 } ),
\label{macd}\end{equation}
where $K$ is a Macdonald function (see  \ref{md}).
In particular,  if $n = 0$ (the Weyl term), \eref{macd} reduces 
to 
$K_{1} (m t )$.
Thus, doing the change of variables \eref{omegatokappa}, we get
\beq\label{EWeyl}
\eqalign{
E_\mathrm{Weyl}(t) &= - \,\frac{  1  }{ 2  }\, \frac{  d }{ d t } 
 \int_{0}^{\infty}  \sigma_\mathrm{Weyl} (\omega) e^{- \omega t 
}\, d\omega  \\
  & = - \,\frac{ 1  }{ 2 \pi }\, \frac{  d }{ d t }  
\int_{0}^{\infty}  \frac{  \sqrt{ \kappa^2 + m^2}   }{ \kappa  } \cdot
  \frac{   \kappa }{ \sqrt{ \kappa^2 + m^2}  } 
e^{- t\,  \sqrt{ \kappa^2 + m^2  } }  \, d\kappa \\
&  =  - \,\frac{ 1  }{ 2 \pi }\,
 \frac{  d }{ d t }  m K_{1} (m t ) 
=\frac{ m^2  }{ 2 \pi }\,
  \left( \frac1{mt}\,K_{1} (m t ) + K_0(mt)\right)}
\eeq
(see \eref{mdderiv}).
Similarly, the periodic term   is
\beq\label{Eper} \eqalign{
E_\mathrm{per} (t) & = -\frac{ 1 }{ \pi} \frac{d}{d t} \sum_{n= 
1}^{\infty} (- 1)^{ n ( l + r) } \int_{0}^{\infty} 
\sigma_\mathrm{per} (\omega ) e^{- \omega t} \, d\omega 
\\
& = -\,\frac{ 1 }{ \pi} \,
\frac{d}{d t} \sum_{n= 1}^{\infty} (- 1)^{ n ( l + r) } \frac{ m 
t }{ \sqrt{ (2 n L)^2 + t^2} } K_{1} (m \sqrt{ (2 n L)^2 + 
t^2} )
.
}\eeq
Finally, 
 the boundary term is
\beq\label{Ebdry}\fl{\eqalign{
E_\mathrm{bdry} (t, x) & = - \,\frac{ (- 1)^{l } } {2 \pi }\, 
\frac{ 
d }{d t } \sum_{n = - \infty}^{\infty} (-1 )^{n (l 
+ r)} \int_{0}^{\infty} \frac{ \omega}{\kappa } \cos( 2 \kappa (x 
+ n L) ) e^{- \omega t }\, d\omega  \\ 
 & = -\, \frac{ (- 1)^{l } } {2 \pi }\, \frac{ d}{d t } \sum_{n = - 
\infty}^{\infty} (-1 )^{n (l + r)} \frac{ m t }{ \sqrt[ ]{ ( 2 (x 
+ n L))^2 + t^2 } }
 K_{1} ( m \, \sqrt{ ( 2 (x + n L))^2 + t^2 } )
.
}}\eeq

\section{Asymptotic behaviors} \label{lims}

\subsection{Small $t$} \label{smallt}
To put the energy expressions \eref{EWeyl}--\eref{Ebdry}
into the usual form for 
renormalization calculations, we need to expand them in power 
(Laurent) series in $t$.
Using \eref{mdpower} one gets
\beq\label{Weylsmall}
\fl{
\eqalign{
E_\mathrm{Weyl} (t) & = - \,\frac{ 1 }{ 2 \pi }\, \frac{ d }{ 
d t} \biggl 
[ \frac{1}{ t}+ \frac{m^2t}{2}   \ln \biggl(\frac{mt}2\biggr )
+\frac{m^2t}4(2\gamma -1) +O\left((m t)^2\right) \biggr ]\\
 & = \frac{1}{ 2 \pi }\, \biggl [ \frac{1 }{ t^2} - \frac{m^2 
}{2 } \ln \biggl ( \frac{m t }{ 2 } \biggr ) - \frac{ m^2 }{ 4 } 
( 2 \gamma +1) \biggr ] + O\left( t \right) .
}}\eeq  
When the derivatives in  \eref{Eper} and \eref{Ebdry} are 
calculated, only one term survives 
 in the limit $t\to0$.
Furthermore, the resulting limits are finite (no negative powers 
or logarithm of $t$):
\beq\label{Eper0} 
E_\mathrm{per} (0) = -\,\frac{ 1 }{ \pi} 
 \sum_{n= 1}^{\infty} (- 1)^{ n ( l + r) } \frac{ m }
{ 2 n L } K_{1} (2  n L m) ,
\eeq
\beq\label{Ebdry0}
E_\mathrm{bdry} (0, x)
  = -\, \frac{ (- 1)^{l } } {2 \pi }  \sum_{n = - 
\infty}^{\infty} (-1 )^{n (l + r)} \frac{ m  }{  2 |x + n L| }
 K_{1} (   2 |x + n L|m ) .
\eeq
In the case \eref{Ebdry}, this argument assumes $x\ne 0$ and 
$x\ne L$, 
and the limit is not uniform in distance from the boundary.
Therefore, we shall need to revisit this case when considering 
the total energy in \sref{totalenergy}.

\subsection{Massless limit} \label{smallm}
The same expansion \eref{Weylsmall} shows that when $m= 0$,
\beq\label{Weylzero}
E_\mathrm{Weyl}(t) = \frac1{2\pi t^2}  \quad(m=0)\,,
\eeq
as expected \cite{SAF}.
(The only interesting fact is that \eref{Weylsmall} includes 
less trivial terms when $m>0$.)
Verifying the massless limits of the infinite 
sums $E_\mathrm{per}(0)$ and 
$E_\mathrm{bdry}(0,x)$ is complicated by the conflict between the 
$m\to0$ and $n\to\infty$ limits in the individual Macdonald 
functions.
However, \eref{Eper0} when $l+r$ is even is a special case of 
\cite[(2.10)]{kir},
a complicated formula from which only one term survives when 
$m=0$:
\beq
E_\mathrm{per}(0) = -\,\frac{\pi}{24L^2}\quad (m=0),
\eeq
the well known one-dimensional Casimir energy.
In exactly the same way, \cite[(2.12)]{kir} gives
$E_\mathrm{per}(0) = +\pi/48L^2$ when $l+r$ is odd 
(one Dirichlet and one Neumann end).

\subsection{Supermassive limit}  \label{largem}
The behavior when $m\to \infty$ is critical for the 
Pauli--Villars analysis.  Using \eref{mdasymp}
one  sees that all the limits are zero:
From \eref{EWeyl} we have
\beq
\lim_{ m \to \infty} E_\mathrm{Weyl} ( t )  =
 \lim_{ m \to \infty}\frac{m^2}{ 2 \pi }
   \sqrt{ \frac{ \pi }{ 2 m t } }\,e^{ - m t } \, = 0 
\end{equation}
when $ t >0$. 
Similarly, the  terms of \eref{Eper0} and \eref{Ebdry0}
(or even \eref{Eper} and \eref{Ebdry}) for fixed $m$ vanish 
sufficiently rapidly with $n$ to make the series converge,
and for fixed $n$ decrease monotonically to $0$ as $m\to\infty$;
therefore, by standard arguments \cite[pp.~3 and~8]{titchmarsh} 
the sum of the series approaches $0$ as $m\to\infty$.
The only exceptions are the terms in \eref{Ebdry0} with $n=0$, 
$x=0$ and with $n=-1$, $x=L$, where the energy density is 
singular, as previously noted.

\section{Total energy}\label{totalenergy}
The energy equals the integral of the energy density over $x$ 
from $0$ to $L$, at least formally.
Departing somewhat from the notation of \cite{SAF},
we denote a total energy by the letter $\overline E$.
$E_\mathrm{Weyl}$ and $E_\mathrm{per}$ are constant in $x$, 
so their energies are obtained  by multiplying by 
$L$ and there is nothing more to be said.

When $l+r$ is even, the boundary formula \eref{Ebdry} yields
\beq
\fl{
\overline{E}_\mathrm{bdry} ( t)  =
 -\, \frac{ (-1)^{l} }{ 2\pi }\, 
\frac{ d}{ d t } \sum_{n = -\infty }^{\infty} \int_{0}^{ L } 
\frac{ m t }{\sqrt[ ]{ 4 (x + n L)^2 + t^2 } }\,
 K_{1} ( m \, \sqrt{ 4 (x + n L)^2 + t^2 } ) \, dx.
  }
\eeq
Making a  change of variables $x' = x + n L$ in each term, we 
have 
\beq\label{bdryenergy}
\eqalign{
\overline E_\mathrm{bdry} (t ) & = - \,\frac{ (-1)^{l} }{2 \pi } 
\,\frac{ d }{ d t } \sum_{n = -\infty }^{\infty} \int_{ L n }^{ 
L (n + 1) } \,
\frac{ m t }{ \sqrt{ 4 x'^2 + t^2 } } K _{1} ( m \, \sqrt{ 
4 x'^2 + t^2 } ) \, dx' \\ 
& =- \,\frac{ (-1)^{l} }{ 2\pi } \frac{ d }{ d t }
 \int_{-\infty }^{ \infty } \frac{ m t }{ 
\sqrt{ 4 x'^2 + t^2 } }\, K _{1} ( m \, \sqrt{ 4 x'^2 +t^2 } 
) \, dx'.
}
\eeq
After another change of variables, $u^2 = 4 x'^2 + t^2$, 
we get with \eref{mdint}
\beq
\eqalign{
\overline{E}_\mathrm{bdry} (t ) & = 
  -\, \frac{ (-1)^{l} }{2 \pi } \,
\frac{d }{ d t } ( m t) \int_{t}^{\infty } 
\frac{ K_{1 } ( m u ) }{ \sqrt{u^2-t^2}} \, du \\
 & = -\, \frac{ 
(-1)^{l} }{4  } \frac{ d }{ d t }   e^{-m t}
 = \frac{ (-1)^{l} }{4}\, m e^{-m t} .
}
\eeq
One can now take $t$ to $0$, obtaining in
 the Dirichlet case ($l = 1$) 
\begin{equation}\label{totbdryenergy}
\overline{E}_\mathrm{bdry} (0) =   -\,\frac{  m  } { 4 } \,  , 
\end{equation}
in agreement with Hays \cite{Hays} and with the conclusion in 
\cite{SAF} that the net boundary energy vanishes in the massless 
case.  It definitely does \emph{not} agree (for any~$m$) with an 
attempt to
integrate $E_\mathrm{bdry}(0,x)$ over the interval  
(that is, to take the cutoff away before integrating),
 which encounters divergences at the endpoints.
For later use note also that
\beq \label{bdryenergyinfty}
\lim_{m\to\infty} \overline{E}_\mathrm{bdry}(t) = 0
\quad\mbox{if (and only if) $t>0$.}
\eeq

When $l+r$ is odd,  $\overline{E}_\mathrm{bdry}(t)$ vanishes for 
an elementary reason: The middle member of 
\eref{bdryenergy} acquires a factor $(-1)^n$, and hence term $n$ 
cancels term $-(n+1)$.

\section{Counting eigenvalues} \label{eigen}
For completeness of the comparison with the massless theory in 
 \cite{SAF}, we look here at the 
 global eigenvalue density, $\rho(\omega)$, 
 and its integral, the counting function  $N(\omega)$.
In \sref{vacuum} we started from the local spectral density, 
$\sigma(x,\omega)$, and integrated in frequency space to get the 
energy density, $E(t, x)\,$; then in \sref{totalenergy} we 
integrated over $x$ to get a total energy.   Here we shall 
perform the integrations in the opposite order.
Looking at the spectral and eigenvalue 
densities is interesting because (unlike most problems) the 
image method determines them exactly, and the eigenvalues are 
known, so that one can directly compare the eigenvalue densities. 
Knowing the eigenvalues, one can then sum over the frequencies,
in one's favorite regularization scheme,
to get the total energy in the traditional Casimir manner, but we 
shall not do that explicitly.

The local spectral density (and hence all the other quantities)
is divided into three qualitatively different parts in the 
defining formula \eref{sigmas}.
The eigenvalue density is thus
\begin{equation}
\rho(\omega) = \int_{0}^{L} \sigma (\omega , x)\, d x = 
\rho_\mathrm{Weyl} (\omega)+ \rho_\mathrm{per} (\omega) + 
\rho_\mathrm{bdry} (\omega),
\end{equation}
where
\begin{equation}
\rho_\mathrm{Weyl} (\omega) = \int_{0}^{L} \sigma_\mathrm{Weyl} 
\, dx 
=\int_{0}^{L} \frac{ \omega }{ \pi \kappa} \, dx = \frac{ L 
\omega }{ \pi \kappa}\, ,
\eeq
and similarly
\beq \rho_\mathrm{per} (\omega ) =
 \frac{2 L\omega }{ \pi \kappa }
 \sum_{n = 1}^{\infty} (- 1 )^{n (l + r) } 
\cos (2 \kappa n L ) ,
\end{equation}
\begin{equation}
\rho_\mathrm{bdry} (\omega ) = \frac{(- 1)^{l} }{ 2 \pi } \sum_{n 
= - \infty}^{\infty} \frac{ (- 1 )^{ n (l + r )} \omega }{ \kappa^2} 
[ \sin ( 2 \kappa L (n + 1 ) ) - \sin ( 2 \kappa L n ) ] ,
\end{equation}
where $\kappa = \sqrt{  \omega^2 - m^2}$.
The eigenvalue counting function equals zero for 
$\omega<m$ and $\int_{0}^{\omega} \rho (\tilde{\omega} ) \, d 
\tilde{\omega }$ for $\omega >m$. 
Therefore, it is (for $\omega >m)$
\beq \label{NWeyl}
N_\mathrm{Weyl}(\omega) = \frac{L}{\pi} \int_{0}^{\kappa} \frac{ 
\omega 
}{\tilde \kappa} \cdot \frac{ \tilde\kappa}{ \omega }\, 
d\tilde\kappa = 
\frac{ L 
\kappa }{\pi} = \frac{ L \, \sqrt[ ]{ \omega^2 - m^2 } }{ \pi}\,,
 \eeq
\beq \label{Nper}
\eqalign{
N_\mathrm{per} (\omega) & = \frac{ 2 L }{\pi } \sum_{ n = 1}^{ 
\infty} 
(- 1 )^{n (l + r) } \int_{0}^{\kappa} \frac{ \sqrt{\tilde 
\kappa^2 + m^2 } }{ \tilde\kappa } \frac{\tilde \kappa }{ 
\sqrt{ \tilde\kappa^2 + m^2 } } 
\cos( 2n L \,\tilde \kappa ) \, d\tilde\kappa \\
& = \frac{ 1}{ \pi} \sum_{ n = 1}^{ 
\infty} \frac{ (- 1 )^{n (l + r) } }{ n } \sin ( 2 n L \, \kappa 
) ,
}
\eeq
and similarly
\beq\label{Nbdry}
\fl{
N_\mathrm{bdry} (\omega)
 = \frac{ (- 1)^{l} }{ 2 \pi} 
\sum_{n = - \infty}^{\infty} (- 1)^{n (l + r)} 
\int_{0}^{\tilde\kappa} 
\frac{ 1 }{ \tilde\kappa } [ \sin ( 2 \tilde\kappa L (n + 1 ) ) - 
\sin ( 2 \tilde\kappa n L ) ] \, d\tilde\kappa .
}
\eeq

The Fourier series in \eref{Nper} can be evaluated 
 as in \cite{SAF} to a sawtooth function, given in 
\eref{saweven}--\eref{sawodd} below.
It is easy to see (as at the end of \sref{totalenergy}) that
$N_\mathrm{bdry}=0$ if $l+r$ is odd (that is, equals $1$).
When $l+r$ is even, we manipulate \eref{Nbdry} to the form
\[
N_\mathrm{bdry} (\omega) =  \frac{ (-1 )^{l} 
}{ \pi} \lim_{n\to\infty}
 \int_{0}^{ \kappa } \frac{ \sin (2 \kappa 
L n ) }{ \kappa} \, d\kappa 
\]
and hence
\beq \label{Nbdryeven}
\fl{
N_\mathrm{bdry} (\omega) =
 \frac{ (- 1)^{l } }{ \pi }\lim_{ n \to \infty}
 \int_{0}^{2nL\kappa} \frac{ \sin \tilde{z } }{ \tilde{z} 
} \, d \tilde{z} 
= \frac{ (-1 )^{l} }{ \pi} \int_{0}^{\infty} 
\frac{ \sin z }{ z}\,dz
 = \frac{ (-1 )^{l } }{ 2}.
}
\eeq

Adding the three counting functions gives
\beq\label{Ntotal}
N (\omega )
  = \frac{L }{ \pi} \, \sqrt{ \omega^2 - m^2} + 
N_\mathrm{per} (\omega )
 + \frac{ (-1 )^{ l } }{ 2 }\,\delta_{l+r,1}
\quad \mbox{(for $\omega>m$)},
\eeq
where
\beq\label{saweven}
N_\mathrm{per} (\omega )  = \frac12 -\frac{L\kappa}{\pi}
\quad\mbox{if $l+r$ is even and %
$\displaystyle 0<\kappa<\frac{\pi}{L}\,$,}
\eeq
\beq\label{sawodd}
N_\mathrm{per}(\omega) = -\,\frac{L\kappa}{\pi}
\quad\mbox{if $l+r$ is odd and %
$\displaystyle -\,\frac{\pi}{2L}<\kappa<\frac{\pi}{2L}\,$;}
\eeq
both functions are extended periodically to all other intervals 
on the positive axis 
of length $\frac{\pi}{L}$ in the variable~$\kappa$.

We now check that  $N(\omega)$ is indeed the number of 
eigenvalues less than or equal to $\omega^2$. 
The true counting function must be $0$ for $\omega<m$
and constant and integer-valued on the interval between two eigenvalues.
Comparing \eref{Ntotal} with \eref{saweven}--\eref{sawodd}, we 
see that $N$ is indeed constant except at the places where 
$N_\mathrm{per}$ has a discontinuity.
At each such point, $N_\mathrm{per}$ jumps from $-\frac12$ to 
$+\frac12$, indicating the addition of one new eigenvalue.
In the odd case, these points occur at 
\beq
\kappa = \left(n-\frac12\right)\frac{\pi}L\,, \quad
\omega^2 = \left(n-\frac12\right)^2\left(\frac{\pi}L\right)^2 
+ m^2 \quad(n=1,2,\ldots)\,, 
\eeq 
and immediately to the right of such a point, $N(\omega)$ 
evaluates to 
\[
\frac{L\kappa}{\pi} +\frac12  =
\left(n-\frac12\right) +\frac12 =n.
\]
That is, the jumps occur at the correct eigenvalues of the mixed
Dirichlet--Neumann problem, and $N$ counts them correctly.
In the even case, the jumps are at numbers of the form
\beq 
\kappa=\frac{n\pi}{L}\,, \quad
\omega^2 = \left(\frac{n\pi}{L}\right)^2 + m^2,
\eeq
and the limit from the right is
\[
N(\omega)=\frac{L\kappa}{\pi} +\frac12  + \frac{(-1)^l}2 =
\cases{n & \mbox{if $l=1$}, \\
n+1  & \mbox{if $l=0$}.}
\]
That is, we get the correct eigenvalues for the Dirichlet and 
Neumann problems, including the extra eigenvalue at 
$n=0$, $\omega=m$, in the Neumann case;
$N_\mathrm{per}$ and $N_\mathrm{bdry}$ conspire beautifully to 
make things come out right at the bottom of the spectrum.

\section{Pressure}
\label{pressure}
Because of the need to deal with spatial derivatives, the 
spectral density $\sigma(\omega,x)$ is not convenient for 
calculating the expectation value of the pressure,
 $p \equiv \langle T_{ 11 } \rangle$.
Therefore, we revert to the formalism of the cylinder kernel,
\beq\label{cylkergen}
\overline{T} (t, \mathbf{r}, \mathbf{r } ' ) = - \sum_{ n = 1 
}^{ \infty } \frac{ 1 }{ \omega_{n } } \phi_{n } ( \mathbf{r } ) 
\phi_{n} ( \mathbf{ r } ' )^{\ast } e^{ - \omega_{n} t } 
\eeq
in terms of  the eigenfrequencies and eigenfunctions 
 of the cavity.
The cylinder kernel for the massive field in infinite space
is \cite[(2.2), (3.1)--(3.2)]{Hays}
\beq\label{cylker0}
\eqalign{
\overline{T}_\infty (t, x,y )&=
-\, \frac{ 1 }{ 2 \pi } \int_{ - 
\infty}^{\infty} \, d\omega \frac{ e^{- i \omega t } }{ 
\sqrt{\omega^2 + m^2 } } e^{ - \sqrt{ \omega^2 + m^{2 }} 
| x - y | } \\
& = -\,\frac{ 1 }{  \pi } K_{0} ( m 
\sqrt{ t^{2 } + (x - y )^2 } ).}
\eeq
The kernel for the Casimir slab is then formed by the same 
construction as in \eref{Greenf}, which again generates Weyl, 
periodic, and boundary terms.
Here and henceforth we confine attention to the pure Dirichlet 
case ($l=r=0$).

 The fundamental formulas for the energy 
density and the pressure in terms of the field are \eref{T00} and 
\eref{T11} in \ref{stress}, from which the formulas for the
vacuum expectation values in terms of $\overline T$ are
\begin{equation} \label{vevE}
E(t,x) = 
-\,\frac{ 1 }{ 2 }\, \frac{ \partial^2 \overline{T} }{ \partial 
t^2 } +\frac{\beta}{2}\biggl ( \frac{ \partial^2 \overline{T} }{ 
\partial x^2 } + \frac{ \partial^2 \overline{T} }{ \partial 
y^{ 2 } } + 2 \frac{ \partial^2 \overline{T} }{ \partial 
x\, \partial y } \biggr ) ,
\end{equation}
\begin{equation} \label{vevp}
p(t,x) = 
\frac{ 1 }{ 8 } \biggl ( \frac{ \partial^2 \overline{T} }{ 
\partial x^2 } + \frac{ \partial^2 \overline{T} }{ \partial 
y^{ 2 } } - 2 \frac{ \partial^2 \overline{T} }{ \partial 
x\, \partial y } \biggr ) ,
\end{equation}
where it is understood, formally, that $y$ is set equal 
to $x$ and $t $ to 0 at the end;
$\beta = \xi-\frac14$ is the curvature (or conformal) coupling 
constant, hitherto assumed to be~$0$.
In arriving at \eref{vevE}--\eref{vevp} several routine 
intermediate steps have been omitted:
Passing from the expectation of the product of two fields to 
$\overline T$ requires inserting a compensating factor 
$\frac12\,$;
field products need to be symmetrized; physical time derivatives 
need to be converted to $t$ derivatives, and in that process 
$\phi_0{}\!^2$ can be converted to 
$-\phi\phi_{00}\,$, so that, in particular, the
$\beta$ term in $p$ turns out to vanish identically.

 The pressure function for the Weyl term, according to 
\eref{vevp} and \eref{cylker0} and the discussion
at the end of \ref{md},  is given by
\begin{equation}
\fl{
\eqalign{
p_\mathrm{Weyl} (t,x,y)   &  = 
 \frac{m }{2 \pi }   \left(\frac{(t+x- y  ) (t-x+ y)
K_1\bigl(m \sqrt{t^2+(x-  y )^2  } \bigr)}
{\left(t^2+(x-  y  )^2   \right)^{3/2} }\right.  \\
&\left.\qquad \qquad  \qquad \qquad \qquad
{}-\frac{m (x-y )^2 
K_0\bigl(m \sqrt{t^2+(x- y   )^2}\bigr)}
{  t^2  +(x-  y )^2 }  \right )  .
}}
\end{equation}
When $y = x$ it simplifies to 
\begin{equation} \label{pWeyl}
p_\mathrm{ Weyl }  (t )  = \frac{m   }{2 \pi  t }\, K_1(m  t 
) . 
\end{equation}

The periodic terms are calculated similarly:
\beq\label{pper}
p_\mathrm{per}(t,x,y) = -\,\frac1\pi 
\frac{d^2}{d x^2}
\sum_{n=1}^\infty K_0\bigl( m\sqrt{(x-y+2nL)^2 +t^2}\bigr).
\eeq
After setting $y=x$ and suppressing the redundant argument, 
we get
\begin{equation} \label{pper1}
\eqalign{
p_\mathrm{ per } (t,x)  &  = - \frac{ m }{   \pi } 
  \sum_{ n = 1 }^{ \infty } 
  \left(\frac{4 L^2 m n^2 K_0
  \left(m \sqrt{4 L^2 n^2+t^2}\right)}{4 L^2 n^2+t^2}
  \right. \\
& \qquad \qquad  \qquad\left. -\,\frac{(t-2 L n) (2 L n+t)
 K_1\left(m \sqrt{4 L^2 n^2+t^2}\right)}
 {\left(4 L^2 n^2+t^2\right)^{3/2}}\right) 
}
\end{equation}
(which actually is independent of $x$).
In fact, here we can immediately set $t=0$, because there is no 
divergence in that limit:
 \begin{equation}\label{pper0}
\eqalign{
 p_\mathrm{ per } (0, x ) &=    
 - \,\frac{ m  }{  \pi }  \sum_{n = 1 }^{ \infty}   
\left(\frac{   K_1(2 m L  n  )}{2   L n  } 
+m K_0(2 m  L  n )\right) \\
 &= \frac{m^2}{\pi} \sum_{n = 1 }^{ \infty}   
K_1'(2mLn)
}
\end{equation}
(by \eref{mdderiv}).
The negative of the periodic pressure \eref{pper0}  correctly 
matches the derivative 
with respect to $L$ of the total periodic energy, which is $L$ times 
quantity \eref{Eper0}:
\beq
\eqalign{
- \,\frac{ d }{ d L } \overline{E}_\mathrm{per} (0) 
&=\frac{d }{d L } \biggl ( \frac{1 }{  \pi } 
\sum_{n = 1 }^{ \infty} 
m^2 \frac{ K_{1 } ( 2 m L n ) }{ 2 m n } \biggr ) \\
&= \frac{ m^2 }{  \pi } \sum_{n = 1 }^{ \infty} 
K_{ 1 } ' (2mLn) = p_\mathrm{per} ( 0,0) \mbox{ or } 
p_\mathrm{per} ( 0,L ).
}
\eeq

The argument at the end of \ref{md} shows that 
 the boundary terms in the  pressure vanish:
\beq \label{bdryp}
 p_\mathrm{bdry}(t,x, y  )= 0 .
\eeq
 This result is analogous to the vanishing of $p_3$ 
 on p.~5 of \cite{EFM}; it reflects the fact that moving the 
boundary does not change the boundary energy.

\section{Conformal correction to the energy} \label{betaterms}
We digress to discuss the ``$\beta$ terms'' in \eref{vevE}.
The same argument from \ref{md}  shows that
the periodic and Weyl $\beta$ terms add to~$0$, 
the sign change on the third term being compensated by 
the replacement of $x+y$ by $x-y$, whereas the boundary 
$\beta$ terms are nonzero,
 in close analogy with \eref{pper}--\eref{pper1}:
\beq\label{K3}
\fl{
\eqalign{
\Delta E^\beta_\mathrm{bdry}(t,x) &=-\, \frac{2\beta m}{\pi}  
\sum_{n=-\infty}^\infty\left ( 
\frac{m 
(2( L 
n+x))^2 K_0\bigl(m 
\sqrt{t^2+(2( L n+x))^2}\bigr)}{(2( L n+x))^2+t^2}\right. 
\\ & 
\left.{}+\frac{(2( L n+x)-t) 
(2 (L n+x)+t)
 K_1\bigl(m \sqrt{t^2+(2( L n+x))^2}\bigr)}
{\left((2( L n+x))^2+t^2\right)^{3/2}} \right ) .
}
}
\eeq
Combining \eref{K3} with \eref{Ebdry0}, we get
\beq \label{Ebdrygenbeta}
\eqalign{
E_\mathrm{bdry}^\beta(0,x) &= E_\mathrm{bdry}(0,x) + 
\Delta E_\mathrm{bdry}^\beta(0,x) \\
&= -\, \frac{1} {\pi } \left(\frac12 + 2\beta\right)
 \sum_{n = - \infty}^{\infty}  \frac{ m  }{  2 |x + n L| }
 K_{1} (2m |x + n L|)
 \\
&\quad{} -\, \frac{2\beta m^2}{\pi}   \sum_{n = - \infty}^{\infty}
K_0(2m |x+ nL|).
}\eeq
If $\beta=-\frac14$ ($\xi=0$), which counts as both conformal and 
minimal coupling in space-time dimension~$2$, then the first 
term in \eref{Ebdrygenbeta} vanishes. The surviving term is less 
singular at the boundary, and it vanishes when $m=0$, as expected 
for a conformally coupled massless field at a flat 
(here, $0$-dimensional) boundary.

\section{Asymptotics of the pressure} 
\label{morepressure}

\subsection{Small $t$ or small $m$} \label{psmall}
From \eref{pWeyl} and \eref{mdpower} we have 
\beq\label{pWeylseries}
p_\mathrm{Weyl}(t)\sim \frac1{2\pi}\left[\frac1{ t^2} 
+\frac{m^2}{2}\,\ln \left(\frac{mt}2\right)
+\frac{m^2}4(2\gamma-1)\right],
\eeq 
and thus
\begin{equation}
p_\mathrm{ Weyl }  ( t )= \frac{   1 }{ 2 \pi t^2 } \quad(m=0) . 
\end{equation}
These formulas are to be compared with \eref{Weylsmall} and 
\eref{Weylzero}.
In fact, we have
\beq\label{Weylsmalldiff}
p_\mathrm{Weyl}(t)-E_\mathrm{Weyl}(t) \sim \frac{m^2}{2\pi}
\left[\ln \left(\frac{mt}2\right)+ \gamma\right],
\eeq
\beq\label{Weylsmallsum}
p_\mathrm{Weyl}(t)+E_\mathrm{Weyl}(t) \sim
\frac1{\pi}\left[\frac1{t^2}-\frac{m^2}4\right].
\eeq
From \eref{Weylsmalldiff} we see that the zero-point 
stress tensor of the massless theory, with the $t$ cutoff,
is traceless ($T^\mu_\mu = -E+p=0$), as befits a conformally 
invariant theory.
On the other hand, \eref{Weylsmallsum} shows that this stress 
tensor does not satisfy the ``principle of virtual work''
(energy-pressure balance)
\cite{Barton,EFM},
$
p = -\,\frac{d\overline{E}}{d L}$
(which is $ - E$ in this case).
A cutoff procedure that respects Lorentz invariance \cite{chris}
must yield a zero-point stress proportional to the metric tensor 
(``dark energy''), replacing \eref{Weylsmallsum} by $0$ but
destroying the tracelessness (unless it makes $p$ and $E$ 
identically~$0$).

For the periodic term we have already taken the cutoff away at
\eref{pper0}, so the only remaining task is to check the massless 
limit in analogy with \sref{smallm}.
In the middle member of \eref{pper0}, 
the first term is the same as \eref{Eper0} (in the Dirichlet 
case), and the second term can be shown to vanish as $m\to0$ by 
\cite[(2.10)]{kir}.
Again $p=E$ in the massless limit.
But in this case we also have the right pressure balance:
\beq
p_\mathrm{per}(0)=+E_\mathrm{per}(0) = -\, \frac{\pi}{24L^2}
= -\,\frac{d}{d L}(LE_\mathrm{per})
\quad(m=0).
\eeq

We have already seen that the boundary pressure vanishes 
identically (as does the $L$ derivative of the boundary energy)
and that the conformally coupled  boundary energy density,
\eref{Ebdrygenbeta}, vanishes when $m=0$,
as does the ``renormalized'' boundary energy, 
\eref{totbdryenergy}.

\subsection{Supermassive limit} \label{plargem}
As $m\to \infty$, the periodic pressure \eref{pper0} 
approaches $0$ because each Macdonald function vanishes 
exponentially rapidly.  The same is true of the Weyl pressure 
\eref{pWeyl} so long as $t\ne0$.  The boundary pressure is 
identically zero.

\section{Applying the Pauli--Villars method} \label{renorm}

In 
sections \ref{smallt} and \ref{psmall} we have shown that the 
stress tensor's expectation value has divergences of orders
$t^{-2}$ and $\,\ln t$.  In dimension $2$ these arise only from 
the zero-point (Weyl) energy, apart from a caveat about a
nonuniform limit at the endpoints in the boundary energy, to 
which we shall return.  The structure is most clearly shown in 
\eref{Weylsmalldiff}--\eref{Weylsmallsum}. 

In these sections, the display of formulas with coincident 
spatial coordinates and small, imaginary time separation is 
purely for calculational and expository convenience; in 
principle, the coordinates are arbitrary.  It may appear that we 
have done a kind of ``point-splitting'' regularization at an 
intermediate step; this perception is wrong.  The spirit of 
Pauli--Villars regularization is to do the subtractions ``at the 
theory level''.  In practice, this means that the subtractions 
involve Green functions \emph{as a whole}, regarded as 
distributions (or, in other words, they involve the operators 
that the Green functions represent).  Thus the potential 
divergences are removed before the issue of evaluating the Green 
functions at coincident arguments ever arises.

Following \ref{hist}, consider the effect of superposing the
stress tensors for several (or many) values of $m$:
\beq\label{PVansatz}
E = \int E[m]\, f(m)\,dm, \quad p = \int p[m]\, f(m)\,dm,
\eeq
where the function or distribution $f$ is independent of~$t$.
If \eref{0int} and \eref{2int} are satisfied,
the terms in \eref{Weylsmallsum} are obliterated;
thus the Weyl part of the vacuum stress satisfies
\beq \label{Weylnonanom}
p_\mathrm{Weyl} = -E_\mathrm{Weyl}
\eeq
(a nontrivial result, in view of \cite{chris} and \cite{EFM}).
We have already observed (\sref{psmall}) that the periodic and 
boundary parts of the stress are also nonanomalous, though the 
relations expressing this health are different from 
\eref{Weylnonanom} because the respective total energies have 
different dependences on~$L$.

If, in addition,
\beq \label{logint}
\int  f(m)\, m^2 \ln{m} \,dm =0,
\eeq
then \eref{Weylsmalldiff} is also obliterated.
If one requires merely that this logarithmic integral be finite, 
as in \eref{2log}, then 
the stress tensor is finite for all $t$ but its Weyl part may be 
a nontrivial multiple of the metric tensor, a two-dimensional 
version of the cosmological constant.

The model as it stands is unlikely to be physically realistic, 
because it contains the effects of unphysical fields with 
negative energies.  Therefore, one studies the effect of taking 
the unphysical auxiliary masses very large, in hopes that their 
effects will become unobservable.  We verified in sections
\ref{largem} and \ref{plargem} that the periodic and boundary 
terms vanish in this limit; only the vacuum stresses of the 
original physical field will survive.  If one can guarantee that 
the integral on the left of \eref{logint} remains finite in the 
limit, then the final theory has no divergences but does have a 
``cosmological'' term with an undetermined coefficient.
If the logarithmic integral is allowed to grow without bound,
to get a finite theory an explicit bare cosmological counterterm 
must be assumed, but the construction is Lorentz-covariant,
unlike \eref{Weylsmallsum}, the result of an ultraviolet cutoff.

The story is different, however, if we look at the total energy.
It is tempting to appeal to \eref{bdryenergyinfty},
but to take the $m$ limit before the $t$ limit would be 
inconsistent with our treatment of the Weyl term.
So, we are stuck with \eref{totbdryenergy}, a boundary energy 
linear in~$m$.  Recall that it arose because of the nonuniform 
limiting behaviors of the boundary stress at the endpoints of the 
interval; in some sense it is \emph{concentrated on the 
endpoints}  and has become independent of the stress in the 
interior, which we have succeeded in regularizing.  Obliterating 
it appears to require yet another constraint on the mass 
distribution~$f$.

In conclusion, we have shown that the Pauli--Villars construction
is mathematically feasible and eliminates the only
 pressure anomaly that arises in dimension~2, the direction 
 dependence of Christensen \cite{chris}.
Physically, whether this maneuver is any more convincing than the 
``analytic'' methods (zeta functions and dimensional 
regularization) is open to debate.  Further philosophical 
discussion probably should wait until an implementation in 
four-dimensional space-time,
where the anomaly of Estrada et al.~\cite{EFM} will be
encountered in the ultraviolet-cutoff theory.

\section*{Acknowledgements}
This research was supported by National Science 
Foundation Grant PHY-0968269,
and by the renewed hospitality of 
the Mathematics Department of Texas A\&M University
toward F.D.M. while some of the work was done.
We thank  Klaus Kirsten and Kim Milton for 
critical remarks on a preliminary presentation of the results.  
{\sl Mathematica\/} was useful at various stages of the work.

\appendix

\section{Varieties of Pauli--Villars regularization}
\label{hist}

Pauli--Villars regularization introduces auxiliary fields
 whose divergences have on balance opposite sign from 
 those of the original, physical fields, so that the total
 predictions of the theory are finite.
 Usually the auxiliary masses are taken very large, so that
 the new fields have no noticeable effects on the finite
 predictions.
 
The original paper of Pauli and Villars \cite{PV}
(which remarks, ``This method has already a long
history,'') deals with quantum electrodynamics in Minkowski
space-time.
Later the method was applied in cosmological contexts 
 \cite{Z,stre,BeDu} and in quantum gravity \cite{ans}.
There are major differences in philosophy and procedure
among these works.

Pauli and Villars distinguish between ``realistic''
and ``formalistic'' regularizations.
In realistic theories the auxiliary masses are assumed to
belong to real (physical) fields, 
whose vacuum energies for some
reason do not all have the same sign;
these masses are kept finite.
In formalistic theories the auxiliary fields are
fictitious, and their masses are sent to infinity at the end
of the calculations.
The realistic approach replaces the original 
theory by a new theory; not surprisingly, the results are not 
unique.  In the formalistic approach after the limit of infinite 
mass, the only surviving ambiguities are those that arise in all 
renormalization schemes and are proportional to the erstwhile 
divergences.  This is our understanding, in the present context, 
of the Pauli--Villars ambiguities recently pointed out in 
\cite{KJ}.  These phenomena are visible below in our two 
treatments of the Zel'dovich regularization of integrals.

Zel'dovich \cite{Z} 
(whose method is followed by Streeruwitz \cite{stre}
without much further discussion of its rationale)
starts from divergent integrals such as
\beq
 \lim_{\Lambda\to\infty}4\pi \int_0^\Lambda 
\sqrt{p^2 + m^2}\,p^2\,dp 
\label{zelddiv}\eeq
and 
postulates a mass distribution function $f(m)$ 
(possibly a finite sum of delta functions)
such that
\beq
\int  f(m)\,dm =\int  f(m)m^2\,dm =\int  f(m)m^4\,dm =0,
\label{zeldfn}\eeq
so that 
\beq
\int dm\, f(m) \int_0^\Lambda p^2\,dp\,  \sqrt{p^2 + m^2}
\label{zeldreg}\eeq
has an asymptotic expansion containing no positive powers
of $\Lambda$.
It may contain terms proportional to
\beq
\int  f(m) m^{2p} \ln(\Lambda m) \,dm \quad (p=0,1,2),
\label{zeldlog}\eeq
but these are actually independent of $\Lambda$ by virtue
of \eref{zeldfn}.
Thus, given a fixed $f$ for which the integrals in
 \eref{zeldlog} with $\Lambda=1$ converge,
 the ultraviolet divergences have been eliminated.
 Because it is not required that the integrals \eref{zeldlog}
 equal $0$, arbitrary renormalization constants appear.
 
 The intention of Zel'dovich was that $f(m)$ represent
 a spectrum of real particles, with negative values of $f$
 arising from fermions.  This theory, therefore, is of the
``realistic'' type; it is a forerunner of supersymmetry.
Zel'dovich's main motivation was to produce a nonzero,
but finite, cosmological constant from the integral
\beq
\int  f(m) m^4 \ln{m} \,dm .
 \label{zeldcosm}\eeq
 
 Note, however, that there is a possibility of creating a
 ``formalistic'' theory by moving the support of the 
 negative part (at least) of $f$
 off to infinity at the end of the argument, provided
 that any integrals like \eref{zeldcosm} that arise
 still converge in this limit.  It is not immediately 
 obvious that this can be done, and especially whether
 the finite limiting values of the renormalization constants
 can be different from $0$.
 It is rather easy to see that the minimal finite sum
 consistent with \eref{zeldfn} will not work:
 Take
 \beq
 f(m) = \delta(m) - \delta(m-m_1) + \delta (m-m_2)
 -\delta(m-m_3),
 \label{3masses}\eeq
where the first term represents the physical field  
(taken here to be massless for simplicity)
and the other three masses are to be taken to $\infty$.
Consider for simplicity a two-dimensional space-time,
so that the only constraints from \eref{zeldfn} and
\eref{zeldlog} that must be satisfied are
\begin{eqnarray}
&\int  f(m)\,dm =0, \label{0int}\\
&\int  f(m)m^2\,dm =0, \label{2int}\\
&\int  f(m)  \ln{m} \,dm <\infty, \label{0log}\\
&\int  f(m) m^2 \ln{m} \,dm < \infty. \label{2log}
\end{eqnarray}
(In \eref{0log} the term with $m=0$ should be omitted.
The precise meaning of \eref{0log} and \eref{2log} is that
the sums remain bounded as the $m_j$ go to infinity.)
It is clear that to satisfy \eref{0int} the total number of 
masses must be even, and then to satisfy \eref{2int} also,
the number must be at least $4$.
Let us first consider the case $m_1=m_3\,$; then by
\eref{3masses},  \eref{2int} becomes
\beq
m_2^2 = 2 m_1^2.
\eeq
Then \eref{2log} states that
\beq  
  m_2^2 \ln m_2 - 2m_1^2 \ln m_1 = 2 m_1^2 \ln \sqrt2
 \eeq
is bounded as $m_1\to \infty$, which is false.
 Now try $m_3=\nu m_1\,$, with $\nu>1\,$:
 After some algebra  one gets
 from \eref{2int} and \eref{2log}
 the same sort of contradiction, unless
 \beq
F(\nu)\equiv (1+\nu^2)\ln(1+\nu^2) - 2\nu^2 \ln \nu =0.
\label{nutry} \eeq
 But $F(1) = 2\ln2>0$ and 
 \beq
 F'(\nu) = 2\nu \ln(1+\nu^2) -2\nu\ln \nu^2 >0,
 \eeq
so splitting the masses can only make the problem worse.

Bernard and Duncan \cite{BeDu} take a formalistic 
approach from the beginning.
They consider a two-dimensional cosmological space-time.
Unlike \cite{Z,stre}, who impose \eref{zeldreg} a posteriori,
they  start with a Lagrangian and
explicitly construct a Fock space.
Their negative-energy fields 
(corresponding to masses $m_1$ and $m_3$ in the foregoing)
are not ordinary fermion fields, but anticommuting
scalar fields producing states with negative metric.
In the infinite-mass limit this sector of the state space 
decouples, leaving a unitary dynamics in a Hilbert space.
This construction apparently requires $m_1=m_3\,$,
so the mass spectrum in \cite{BeDu} is the same as
\eref{3masses}, except that they allow the physical field 
to have a mass, $\mu\,$; then $m_2^2= 2M^2 -\mu^2$ and
(in effect) $m_1^2=m_3^2 = M^2$.  Thus \eref{0int} and
\eref{2int} are satisfied.
It turns out that \eref{0log} is unnecessary because of
the triviality of two-dimensional gravity.
But \eref{2log} is not satisfied in the limit;
instead, Bernard and Duncan explicitly introduce a 
cosmological counterterm to cancel this divergence.
They remark that the analogous construction in dimension~4
would require seven regulator  fields 
(counting the complex anticommuting ones twice) and
(as usual in four-dimensional gravity)  four counterterms.
(Without the constraint that both masses associated with
an anticommuting field be the same, 
the three regulator masses in \eref{3masses}
would be enough to satisfy \eref{zeldfn}.)

In the Bernard--Duncan approach, then, the Pauli--Villars
construction does not, by itself, remove infinities, but
it does achieve covariance:  Divergences in the limit
of large cutoff, $\Lambda$, are replaced by divergences
in the limit of large~$M$.  The regulated (finite-$M$)
expressions are free of the direction dependence 
\cite{chris}
and resulting pressure anomalies \cite{EFM} 
characteristic of point-splitting calculations of the
stress tensor.

Anselmi \cite{ans} takes the further step of cancelling 
the large-$M$ divergences by adding still more regulator
fields.  He requires that the logarithmic
sums \eref{zeldlog} vanish.  He inserts the regulator
fields into a path integral in a nonstandard way, which
permits (in effect) spectra like \eref{3masses} with
coefficients not necessarily equal to $\pm 1$.  This
additional freedom allows the conditions to be satisfied
by solving a linear system for those coefficients, instead
of the nonlinear system for the masses;
this significantly simplifies the study of the existence 
question.
The result is that, with enough regulator fields, a
formalistic formulation without counterterms is achieved.
(Nevertheless, because of the need to modify \eref{zeldlog}
for $p=0$ when the physical field is massless, 
the logarithmic divergences inevitably
result in two arbitrary renormalized coupling constants
in the final equation of motion of the gravitational field.
In the present paper this complication does not concern us.)

\section{Calculus with Macdonald functions}\label{md}

In \cite{GR} or any similar reference one finds the
approximations
\beq\label{mdpower}
K_1(z) = \frac1z  +\frac z2\ln{\frac 
z2}+\left(\gamma-\frac12\right)\frac z2+O(z^3\ln z)
\eeq
for small $z$ and
\beq\label{mdasymp}
K_\nu(z) \sim \sqrt{\frac\pi{2z}} \,e^{-z}
\eeq
for large $z$.

Derivatives can be eliminated by \cite[(8.486.12)]{GR}
\beq \label{mdderiv}
K_0'(z) = -K_1(z), \qquad
K_1'(z) = -\,\frac1z K_1(z) - K_0(z).
\eeq

The integral
\beq \label{mdint}
\int_t^\infty \frac {K_1(mu)}{\sqrt{u^2-t^2}}\,du = 
\frac {\pi e^{-mt}}{2mt}
\eeq
does not appear in \cite{GR} but is known to 
{\sl Mathematica} and
can be deduced from \cite[(6.596.3)]{GR}.

In \sref{pressure} and \sref{betaterms} we repeatedly 
encounter second derivatives of
\beq\label{mdbasic}
K_0\bigl(m\sqrt{(x\pm y+2nL)^2+t^2}\bigr).
\eeq
The results are simplified by \eref{mdderiv} and
further simplified by cancellations and combinations: 
Looking, for example, at \eref{vevp}, one can show that
the first two terms are equal and their sum is equal to
the third term up to sign.  Thus the whole expression
vanishes if the variable sign in \eref{mdbasic} is~$+$
and equals 4 times the first term if that sign is~$-$.
For the $\beta$ terms in \eref{vevE} the role of the sign
is precisely the reverse.

\section{The stress tensor in dimension 2}\label{stress}

The general form of the scalar stress tensor, defined by 
variation of the gravitational Lagrangian with respect to the 
metric, is given (in the sign convention where $g_{00}<0$)
in \cite{DeWitt,chris,BDavies}.
After specializing to flat space (and glossing over operator 
symmetrizations), it is
\beq \label{genstress}
\fl{
\eqalign{
T_{\mu\nu} &= (1-2\xi)\phi_\mu \phi_\nu 
+ (2\xi- {\textstyle{1\over 2}})g_{\mu\nu} \phi_\sigma 
\phi^\sigma 
-2\xi\phi\phi_{\mu\nu}
+2\xi g_{\mu\nu} \phi\phi^\sigma_\sigma
-{\textstyle{1\over2}} m^2 g_{\mu\nu}\phi^2 \\
& = {\textstyle{1\over 2}}[\phi_\mu \phi_\nu 
-\phi\phi_{\mu\nu} 
+ g_{\mu\nu} \phi\phi^\sigma_\sigma
- m^2 g_{\mu\nu}\phi^2]
+2\beta[-\phi_\mu \phi_\nu 
+ g_{\mu\nu} \phi_\sigma \phi^\sigma 
-\phi\phi_{\mu\nu}
+ g_{\mu\nu} \phi\phi^\sigma_\sigma],
}}\eeq
where $\xi\equiv\beta+{1\over 4}$ is the curvature coupling 
constant  (and indices on $\phi$ denote derivatives).
Using the equation of motion, 
$\phi_\sigma^\sigma = m^2 \phi$,
 to rewrite the first term (but not the second),
one arrives at
\beq \label{mygenten}
T_{\mu\nu} = {\textstyle{1\over 2}}[\phi_\mu \phi_\nu 
-\phi\phi_{\mu\nu}] 
+2\beta[-\phi_\mu \phi_\nu 
+ g_{\mu\nu} \phi_\sigma \phi^\sigma 
-\phi\phi_{\mu\nu}
+ g_{\mu\nu} \phi\phi^\sigma_\sigma].
\eeq
The advantages of this form are (a) the mass (more generally, a 
scalar potential \cite{fall4}) does not appear at all,
(b) the first term of 
$T_{\mu\mu}$ contains only $\mu$ derivatives, and (c)
the $\beta$  term is manifestly a total derivative.
Specializing henceforth to dimension $1+1$, we have
\beq \label{T00}
T_{00}= {\textstyle{1\over 2}}[\phi_0^2 - \phi \phi_{00}] 
-2\beta[\phi_1^2 +\phi\phi_{11}],
\eeq
\beq \label{T11}
T_{11}= {\textstyle{1\over 2}}[\phi_1^2 - \phi \phi_{11}] 
-2\beta[\phi_0^2 +\phi\phi_{00}]. 
\eeq


\Bibliography{00} \frenchspacing

\bibitem{Milton}  Milton K A 2001 \emph{The Casimir Effect: 
Physical Manifestations of Zero-Point Energy} Singapore:World 
Scientific.

\bibitem{BKMM} Bordag M, Klimchitskaya G L, Mohideen U and
Mostepanenko V M 2009
\emph{Advances in the Casimir Effect} Oxford:Clarendon Press.

\bibitem{LosAlamos}  Dalvit D, Milonni P, Roberts D and da Rosa F
(eds.)\ 2011
\emph{Casimir Physics} (Lecture Notes in Physics 834)
Berlin:Springer.



\bibitem{Hays}
 Hays P 1979
 Vacuum fluctuations of a confined massive field in 
two dimensions
\emph{Ann. Phys.} \textbf{121} 32--46

\bibitem{SAF}  Fulling S A 2007
 Vacuum energy as spectral geometry
\emph{SIGMA} \textbf{3} 094 [arXiv:0706.2831]

\bibitem{PV} Pauli W and Villars F 1949
On the invariant regularization in relativistic
quantum theory
\emph{Rev. Mod. Phys.} 
\textbf{21} 434--444

\bibitem{Z} Zel'dovich Ya B 1968
The cosmological constant and the theory of elementary
particles
\emph{Sov. Phys. Usp.} \textbf{11} 381--393
[\emph{Usp. Fiz. Nauk} \textbf{95} 209--230]

\bibitem{stre} Streeruwitz E 1975
 Vacuum fluctuations of a quantized 
scalar field in a Robertson-Walker universe
\emph{Phys. Rev. D} \textbf{11} 
 3378--3383



\bibitem{BeDu} Bernard C and  Duncan A 1977
 Regularization and 
renormalization of quantum field theory in curved space-time
\emph{Ann. Phys.} \textbf{107}  201--221

\bibitem{ans} Anselmi D 1993
Covariant Pauli--Villars regularization of quantum gravity
at the one-loop order
\emph{Phys. Rev. D} \textbf{48} 5751--5763


\bibitem{EFM}
 Estrada R, Fulling S A, and  Mera F D 2012
 Surface vacuum energy in cutoff models: 
Pressure anomaly and distributional gravitational limit
 \emph{J. Phys. A} \textbf{45} 455402

\bibitem{BH}
Bender C M and  Hays P 1976
Zero-point energy of fields in a finite volume
 \emph{Phys. Rev. D} \textbf{14} 2622--2632

\bibitem{GR}  Gradshteyn I S and Ryzhik I  M 1965
\emph{Table of Integrals, Series and Products}  New York:Academic 
Press 


\bibitem{kir} Kirsten K 1992
Connections between Kelvin functions and zeta functions with 
applications
\emph{J. Phys. A} \textbf{25} 6297--6305

\bibitem{titchmarsh}Titchmarsh E. C. 1939 \emph{The Theory of 
Functions} (2nd ed) Oxford:Oxford University Press




\bibitem{Barton} Barton G 2004
Casimir's spheres near the Coulomb limit:
energy density, pressures and radiative effects
\emph{J. Phys. A} \textbf{37} 3725--3741







\bibitem{chris}  Christensen S M 1976
 Vacuum expectation value 
of the stress tensor in an arbitrary curved background: The 
covariant point-separation method
\emph{Phys. Rev. D} \textbf{14} 2490--2501

\bibitem{KJ} Kleiss R H P and Janssen T 2014
Ambiguities in Pauli--Villars regularization
\texttt{arXiv:1405.1536}.



\bibitem{DeWitt} DeWitt B S 1975
 Quantum field theory in curved spacetime
\emph{Phys. Rep.} \textbf{19}  295-357

\bibitem{BDavies}
 Birrell  N D and  Davies P C W 1982
 \emph{Quantum Fields in Curved Space}
Cambridge:Cambridge University Press.

\bibitem{fall4} Milton K A, Shajesh K V, Fulling S A, and 
Parashar P 2014
How does Casimir energy fall? IV 
\emph{Phys. Rev. D} \textbf{89} 064027


\end{thebibliography}
\end{document}